%% file: main.tex
\documentclass[conference]{IEEEtran}

\usepackage{booktabs}
\usepackage{graphicx}
\usepackage{xcolor}
\usepackage{amsmath}
\usepackage{enumitem}
\usepackage{url}
\usepackage{cite}

\usepackage{eso-pic}
\AddToShipoutPictureFG*{%
  \AtPageLowerLeft{%
    \raisebox{12pt}[0pt][0pt]{%
      \makebox[\paperwidth]{%
        \parbox{7.5in}{\fontsize{6}{7}\selectfont
          Author accepted manuscript. Published in IEEE CLOUD 2026, pp. 414--421. DOI: \url{https://doi.org/10.1109/CLOUD72782.2026.00054}\\
          \copyright\ 2026 IEEE. Personal use of this material is permitted. Permission from IEEE must be obtained for all other uses, in any current or future media, including reprinting/republishing this material for advertising or promotional purposes, creating new collective works, for resale or redistribution to servers or lists, or reuse of any copyrighted component of this work in other works.
        }%
      }%
    }%
  }%
}

\begin{document}

\title{InferScale: Kubernetes-Native Predictive Autoscaling for SLO-Aware AI Inference}

\author{\IEEEauthorblockN{Wen Huang}
\IEEEauthorblockA{Independent Researcher\\
alexwhuang@outlook.com}}

\maketitle

\begin{abstract}
Kubernetes Horizontal Pod Autoscaler (HPA) is widely used for inference services, but reactive scaling is often too slow when pods take tens of seconds to become ready. The problem gets worse in shared clusters, where independent HPAs compete for the same resource budget and can hurt different tenants from run to run. We present \textbf{InferScale}, a Kubernetes-native autoscaler that combines inference-aware signals, short-horizon demand prediction, SLO-state control, fast-path spike handling, and tenant-priority headroom without replacing the default scheduler.

We evaluate InferScale on Azure Kubernetes Service in two settings. On a real ResNet-50 ONNX workload with 21-second cold start, repeated ramp-burst runs improve SLO compliance from 47.8\% $\pm$ 1.5\% under a stronger external-metric HPA to 65.3\% $\pm$ 9.0\% under InferScale. In repeated controlled multi-tenant contention experiments, InferScale cuts Gold-tier SLO standard deviation from 16.1 to 1.2 percentage points and makes contention outcomes follow the priority policy more closely. Repeated single-service ablations show 6--8 percentage point degradation, with an exploratory priority ablation showing a 27-point drop under contention. The main takeaway is that, in startup-sensitive inference serving, the practical value of prediction lies less in precise forecasting than in moving the controller early enough to offset cold-start delay.
\end{abstract}

\begin{IEEEkeywords}
Kubernetes, autoscaling, AI inference, SLO-aware systems, predictive scaling, multi-tenant cloud systems
\end{IEEEkeywords}

\section{Introduction}

AI inference is now a normal Kubernetes workload. Teams deploy model servers as standard Deployments, expose them over HTTP, and expect the platform to scale them the same way it scales other services. That is convenient operationally, but inference serving has one awkward property: new pods are often expensive to start.

A serving pod may need to download model artifacts, initialize a runtime, allocate memory, and run warmup requests before it is actually useful. In our ONNX Runtime setup, a ResNet-50 pod takes about 21 seconds to become ready even on a modest 500m-CPU allocation. HPA reacts only after a metric crosses a threshold~\cite{k8shpa}, and the rest of the path---scrape, reconcile, create, ready---adds more delay. In our AKS deployment, that whole loop takes roughly 60--90 seconds for real model serving. For a short burst, that is simply too late.

We first expected a stronger reactive baseline to narrow the gap. It helped, but not enough. In a motivation experiment with an external-metric HPA, the autoscaler still violated the 200\,ms P95 SLO for 60 of 61 burst seconds. Static provisioning held the SLO, but only by keeping extra replicas around when they were not needed. That gives the basic tradeoff for this paper: react and miss, or provision and waste.

Shared clusters make the problem less predictable. When several inference services share the same node pool, independent HPAs can hurt different tenants in different runs even under the same workload and budget. Large production systems have long faced the same resource-sharing tension at cluster scale~\cite{borg}. For a cloud platform team, that is a bad property. The problem is not just average latency; it is that the policy outcome is hard to reason about.

\textbf{InferScale} is our attempt to fix that without changing the deployment model. It stays in the standard Kubernetes control plane and does not replace the scheduler. The controller uses inference-aware signals, short-horizon demand prediction, SLO-state control, fast-path spike handling, and tenant-priority headroom to move earlier when startup delay is the real bottleneck. The contribution is not a new forecasting algorithm in isolation; it is the integration of early demand signal, SLO-state response, and priority-aware headroom into a Kubernetes-native control loop and the evaluation of where that integration helps.

The main contributions are:
\begin{itemize}[leftmargin=1.2em,nosep]
    \item We show, with both real-model and controlled evidence, that reactive HPA is too slow for bursty inference workloads when pod startup is non-trivial.
    \item We present InferScale, a Kubernetes-native predictive autoscaler that combines short-horizon demand estimation, SLO-state control, fast-path spike handling, and tenant-aware guardrails within the standard control plane.
    \item We evaluate InferScale on AKS and show that repeated real-model ramp-burst runs still improve SLO compliance over both CPU HPA and a stronger external-metric HPA, while also providing more consistent protection for high-priority tenants under repeated shared-cluster contention.
    \item Ablation shows that under startup-sensitive bursts, early trend exposure matters at least as much as the forecasting component itself. Removing individual components degrades repeated single-service SLO compliance by 6--8 percentage points, while an exploratory priority ablation shows a 27-point drop under contention.
\end{itemize}

\section{Background and Motivation}

\subsection{Why reactive autoscaling struggles with inference}

HPA works well when pods start quickly and the scaling metric tracks load tightly. Inference weakens both assumptions. Model-serving pods download artifacts, initialize runtimes, and run warmup inferences before becoming ready---21 seconds in our ResNet-50 setup. Meanwhile the user-visible objective is tail latency, not the indirect signal that the autoscaler watches.

The resulting reaction path is long: scrape metrics, observe a threshold crossing, create replicas, and wait for readiness. In our AKS deployment, total response time reaches 60--90 seconds. A short burst can therefore cause sustained SLO violations before new capacity arrives.

\subsection{Motivation experiment}

\begin{figure}[t]
    \centering
    \includegraphics[width=0.95\linewidth]{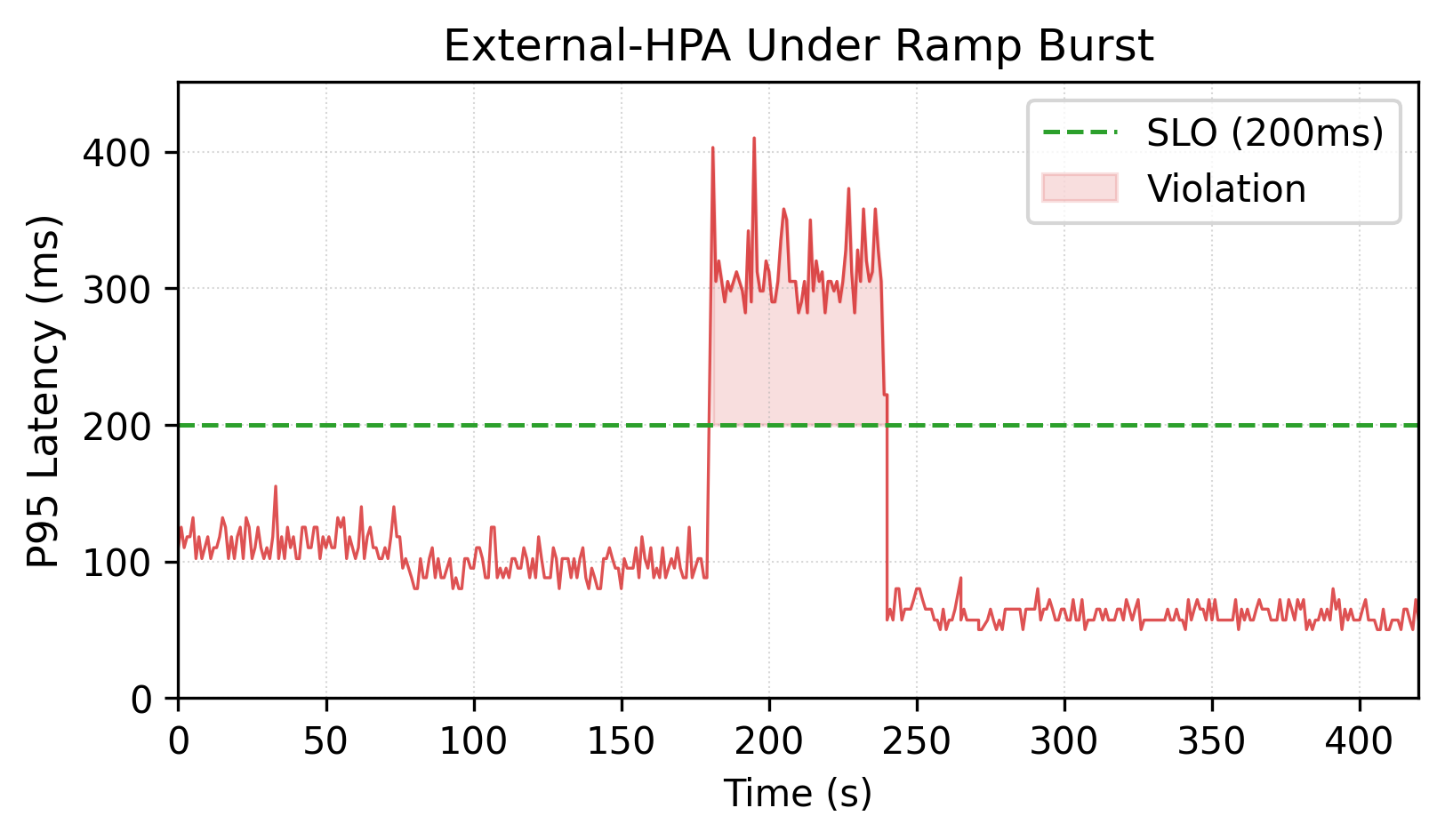}
    \caption{Motivation experiment. External-metric HPA violates the SLO for nearly the entire burst because reactive decisions arrive after the useful window.}
    \label{fig:motivation}
\end{figure}

We ran a controlled motivation experiment on AKS: a burst from 20 to 120\,QPS against an inference service with a 45-second startup delay. This setup is separate from the later real-model ResNet-50 evaluation and is used only to illustrate why stronger reactive baselines can still miss short bursts. As Figure~\ref{fig:motivation} shows, CPU HPA barely reacted. External-metric HPA reacted, but too late---it violated the 200\,ms P95 target for 60 of 61 burst seconds. Static provisioning with six replicas held the SLO at the cost of persistent over-provisioning.

\subsection{Why shared-cluster pressure matters}

In production clusters, multiple inference services share a node pool. When resources become scarce, independent reactive autoscalers can interact badly: one tenant scales first and consumes most of the budget, starving others. We observed exactly this---HPA produced unstable per-tenant outcomes across repeated trials with identical workloads and budgets.

Beyond simply scaling faster, a useful autoscaler should also behave predictably under shared-cluster pressure.

\section{InferScale Design}

InferScale is a Kubernetes operator that manages a custom resource (\texttt{InferenceAutoscaler}) specifying the target Deployment, SLO, replica bounds, forecast window, and tenant class. The controller queries Prometheus every 5 seconds, computes a desired replica count, and patches the Deployment through the Kubernetes API.

\subsection{Architecture}

Figure~\ref{fig:arch} shows the architecture. InferScale follows the standard Kubernetes operator pattern. An \texttt{InferenceAutoscaler} custom resource specifies the target Deployment together with policy inputs such as SLO, tenant class, replica bounds, forecast window, and cooldown. The controller then executes a reconcile loop every 5 seconds.

Each reconciliation cycle has four stages:
\begin{enumerate}[leftmargin=1.2em,nosep]
    \item \textbf{Collect} request rate, queue depth, P95 latency, and ready-replica count from Prometheus.
    \item \textbf{Predict} near-future demand using a short-horizon forecasting model.
    \item \textbf{Decide} a target replica count from predicted demand, estimated capacity, and current SLO state.
    \item \textbf{Guard} the decision with bounded scaling and cooldown-based scale-down before patching the Deployment replica count through the Kubernetes API.
\end{enumerate}

InferScale does not modify the serving runtime or replace the default scheduler. Existing Kubernetes services can adopt it by exporting the required metrics and attaching an autoscaler custom resource.

In the reported experiments, the controller uses a 60-second forecast window.

This design still inherits the latency of scraped metrics. Prometheus scraping, adapter propagation, and controller reconciliation mean InferScale cannot observe a burst at the exact instant it begins. The design therefore should not be read as eliminating monitoring delay. Instead, it tries to use the first available rising signal more effectively than a threshold-only reactive controller. Once the signal appears, the predictor and fast path move the replica target before the service has already spent a full startup interval in violation.

\input{fig_inferscale_arch.tex}

\subsection{Inference-aware signals and demand estimation}

InferScale uses a small, portable signal set---QPS, queue depth, P95 latency, and ready replicas---that is easy to expose from common model-serving stacks. It then estimates near-future request rate using a lightweight Holt-style forecasting model~\cite{holt}:
\begin{align}
\ell_t &= \alpha x_t + (1-\alpha)(\ell_{t-1} + b_{t-1}) \\
 b_t &= \beta (\ell_t - \ell_{t-1}) + (1-\beta)b_{t-1}, \\
\hat{x}_{t+k} &= \ell_t + kb_t.
\end{align}
We use the predictor for a short horizon rather than attempting long-range workload forecasting. In this setting, the purpose of prediction is not perfect accuracy; it is to expose enough trend signal for the controller to move before the burst fully lands.

We kept the predictor simple on purpose. A heavier forecasting model might improve fit on some traces, but it would not change the basic control problem this paper focuses on: once startup takes tens of seconds, the controller mainly needs a reliable early signal, not a sophisticated long-range forecast. That is why we spend more of the design budget on SLO-state control, fast-path handling, and priority-aware decisions than on predictor complexity.

Baseline replica demand is computed from the larger of current and predicted QPS divided by estimated per-replica capacity. Capacity is tracked with an EWMA over healthy observations only. Under overload, observed throughput drops due to queueing and readiness lag, not because true capacity changed; updating from those samples would underestimate capacity and amplify oscillation.

\subsection{SLO-state control and fast path}

The controller evaluates the ratio between observed P95 latency and the configured SLO target and classifies the service as Healthy, Pressure, or Violation. In the reported experiments, Pressure begins at a P95/SLO ratio of 0.7 and Violation begins at 1.0. Headroom increases with severity: Pressure triggers more aggressive scaling than Healthy, while Violation adds an emergency step.

A separate fast path handles rapid load increases. In the shared-cluster experiments, fast-path spike detection is triggered when the relative QPS growth over the recent control window exceeds 0.5, allowing the controller to add headroom before the smoothed trend fully converges. This matters near the rising edge of a burst, where a predictor alone can still lag.

\subsection{Tenant-aware control and guardrails}

InferScale supports tenant classes such as Gold, Silver, and Bronze. In the reported multi-tenant experiments, each custom resource specifies a tenant class, a per-tenant P95 SLO, a 60-second forecast window, and a 120-second scale-down cooldown. Under scarcity, the controller uses tenant class to differentiate headroom so that independent scaling decisions do not collapse into winner-takes-all behavior.

This mechanism is related to keeping spare capacity for important tenants, but it is applied through the autoscaling policy rather than through a separate warm-pool manager. Static extra capacity is a strong operational baseline for burst protection, especially for high-priority tenants, but it also pays idle cost continuously. InferScale instead adds headroom when the observed and predicted state justify it. Warm pools and pre-provisioned standby pods are therefore complementary: they would reduce the physical cold-start bound that any autoscaler faces, while InferScale decides when that capacity should be requested.

To remain practical under noisy scraped metrics, InferScale combines bounded per-cycle scaling with cooldown-based scale-down. The exact thresholds are configuration parameters; in the experiments reported here, the scale-down cooldown is set to 120 seconds. The goal of these guardrails is not sophistication, but stability.

\section{Evaluation}

We evaluate three questions:
\begin{itemize}[leftmargin=1.2em,nosep]
    \item \textbf{Q1.} Does InferScale outperform reactive baselines on real model serving?
    \item \textbf{Q2.} Does it produce more predictable outcomes under multi-tenant contention?
    \item \textbf{Q3.} Which components matter most?
\end{itemize}

\subsection{Setup}

\textbf{Cluster.} All experiments run on AKS in West US 2 with two \texttt{D2s\_v3} system nodes and three \texttt{D4s\_v3} user nodes.

\textbf{Real-model workload.} Each pod requests 500m CPU and 512\,MiB (Guaranteed QoS). The container downloads ResNet-50 from Azure Blob, loads it into ONNX Runtime, runs 60 warmup inferences, and then reports ready. Cold start is approximately 21\,s. Effective capacity is approximately 2--3\,QPS per pod at the configured 1500\,ms P95 target.

\textbf{Multi-tenant workload.} Three tenants (Gold/Silver/Bronze) share the cluster using a controlled service with configurable latency and startup delay. This isolates allocation behavior and enables repeatable trials.

\textbf{Comparison fairness.} All single-service methods use the same pod resource requests, the same cluster budget, and the same min/max replica bounds. The multi-tenant experiments likewise use the same cluster, per-tenant resource requests, and replica bounds across methods. Differences therefore come from control decisions rather than a looser resource budget. Unless otherwise noted, we use a 60-second forecast window, a 120-second scale-down cooldown, and fixed forecasting and guardrail settings across InferScale runs.

\textbf{Baselines.} Real-model experiments compare InferScale against both CPU HPA (target 50\% utilization), the operational baseline many deployments start with, and a stronger external-metric HPA based on queue depth. The external-metric HPA uses the same exported queue-depth metric as InferScale, a target value of 8 queued requests, and the same single-service replica bounds (1--10). In the multi-tenant experiments, each tenant has its own external-metric HPA with the same queue target; Gold is allowed up to 10 replicas, while Silver and Bronze are allowed up to 8, matching the InferScale custom resources. We did not add extra smoothing or forecasting to this baseline, because the point of comparison is a well-tuned reactive controller over inference-specific metrics. The motivation experiment additionally includes static provisioning, showing that reactive control remains limited by startup delay while static capacity pays an idle-cost penalty.

We separate the single-service and multi-tenant studies because they answer different questions. The real-model study tests whether the controller helps under realistic startup and serving costs. The controlled multi-tenant study asks whether the allocation outcome becomes easier to predict under shared-cluster pressure. This is still not a full real-model multi-tenant validation.

The baseline choice follows the same logic. The motivation experiment includes external-metric HPA and static provisioning to show that even a stronger reactive controller still misses short bursts. Q1 includes both CPU HPA and external-metric HPA so the paper is not reduced to a CPU-HPA comparison. Q2 keeps the stronger reactive baseline as well, so the contention result is not attributed to a weak comparator.

The paper also uses three different evidence levels. CPU HPA is kept as an operational reference even though severe overload reduces usable coverage. The stronger reactive baseline is repeated where the main concern is comparator stability. Ramp-burst single-service results therefore rely most on repeated comparisons, whereas diurnal and priority-ablation results are interpreted more conservatively.

\textbf{Reproducibility.} The implementation and experiment setup are configuration-driven. The controller behavior is defined through Kubernetes manifests and autoscaler custom resources, while the reported runs are generated by the same load scripts, Prometheus queries, and deployment templates across baselines. The artifact set includes controller and HPA manifests, repeated-run directories with per-run logs and replica traces, and the figure-generation scripts used in the paper. We do not claim push-button reproduction on arbitrary clusters, but the pipeline from configuration to metrics to reported tables is explicit enough to audit.

Different parts of the evaluation have different noise profiles. Burst-heavy single-service runs are sensitive to startup timing and control cadence, so repeated comparison matters there. By contrast, the Gold-tier multi-tenant result remains stable across repeated contention runs, while the lower-tier story is better understood as a policy tradeoff than as a low-variance performance claim.

\subsection{Q1: Real-model validation}

\begin{table}[t]
\centering
\caption{Real-model validation (ResNet-50, ONNX Runtime, P95 $<$ 1500\,ms). Latency in ms. CPU-HPA is shown as the operational single-run reference because loadgen timed out under overload. Ramp External-HPA and InferScale results are repeated means over three runs. Diurnal External-HPA and InferScale results are representative single-run values from the same setup.}
\label{tab:realmodel}
\scriptsize
\begin{tabular}{@{}lcccc@{}}
\toprule
\textbf{Experiment} & \textbf{SLO\%} & \textbf{P95 med} & \textbf{Avg Repl.} & \textbf{Max Repl.} \\
\midrule
ramp-cpu-hpa & 30.6 & 3494 & 1.5 & 3 \\
ramp-ext-hpa & 47.8 $\pm$ 1.5 & 1650 & 2.1 & 7 \\
\textbf{ramp-inferscale} & \textbf{65.3 $\pm$ 9.0} & \textbf{1097} & 5.5 & 7 \\
\midrule
diurnal-cpu-hpa & 58.6 & 1268 & 2.0 & 3 \\
diurnal-ext-hpa & 74.9 & 618 & 5.7 & 10 \\
\textbf{diurnal-inferscale} & \textbf{80.3} & \textbf{572} & 5.1 & 7 \\
\bottomrule
\end{tabular}
\end{table}

Table~\ref{tab:realmodel} shows the updated real-model comparison. External-metric HPA is a substantial upgrade over CPU HPA, and repeated ramp-burst runs still favor InferScale: 65.3\% $\pm$ 9.0\% versus 47.8\% $\pm$ 1.5\%. We repeated the stronger reactive comparator and the main method in this ramp setting because the main concern was whether the apparent gain came from baseline or controller instability rather than from the design itself.

The repeated ramp results also sharpen the paper's scope. Under the ramp burst, the gap remains large. Under the diurnal pattern, the two methods are much closer (74.9\% vs 80.3\%), so we do not treat that regime as a headline win.

The contrast between the two workloads is itself informative. In the ramp case, the system spends most of the critical interval deciding and waiting: by the time a reactive controller has enough evidence to scale, much of the useful window is already gone. In the diurnal case, load changes more gradually, so a stronger reactive baseline has time to catch up and the predictive advantage shrinks.

The repeated ramp runs are informative in another way as well. InferScale has noticeably higher variance than the stronger reactive baseline. That is not necessarily a bug. External HPA is more deterministic because it waits for a clearer signal and reacts later. InferScale moves earlier, so it is more exposed to timing details in scrape cadence and control alignment. What matters is that the separation between the two systems still holds at the regime level: the worst InferScale ramp run remains above the best repeated external-HPA run.

If an operator only cared about minimizing run-to-run variance, the stronger reactive baseline would look attractive. But that stability comes from reacting late and giving up the early part of the burst almost entirely. InferScale accepts more timing sensitivity in exchange for consistently operating in a better part of the latency curve during the interval that actually matters.

A closer look at the time series explains the difference. In the ramp run, the reactive baselines remain underprovisioned through the 6\,QPS and 12\,QPS phases, then begin scaling only after the service is already in sustained distress. InferScale, by contrast, reaches six replicas around the start of the heavy-load phase and seven replicas near the peak. This does not mean prediction is perfect. It means the controller moves early enough to keep the service closer to its operating region.

\begin{figure}[t]
    \centering
    \includegraphics[width=0.96\linewidth]{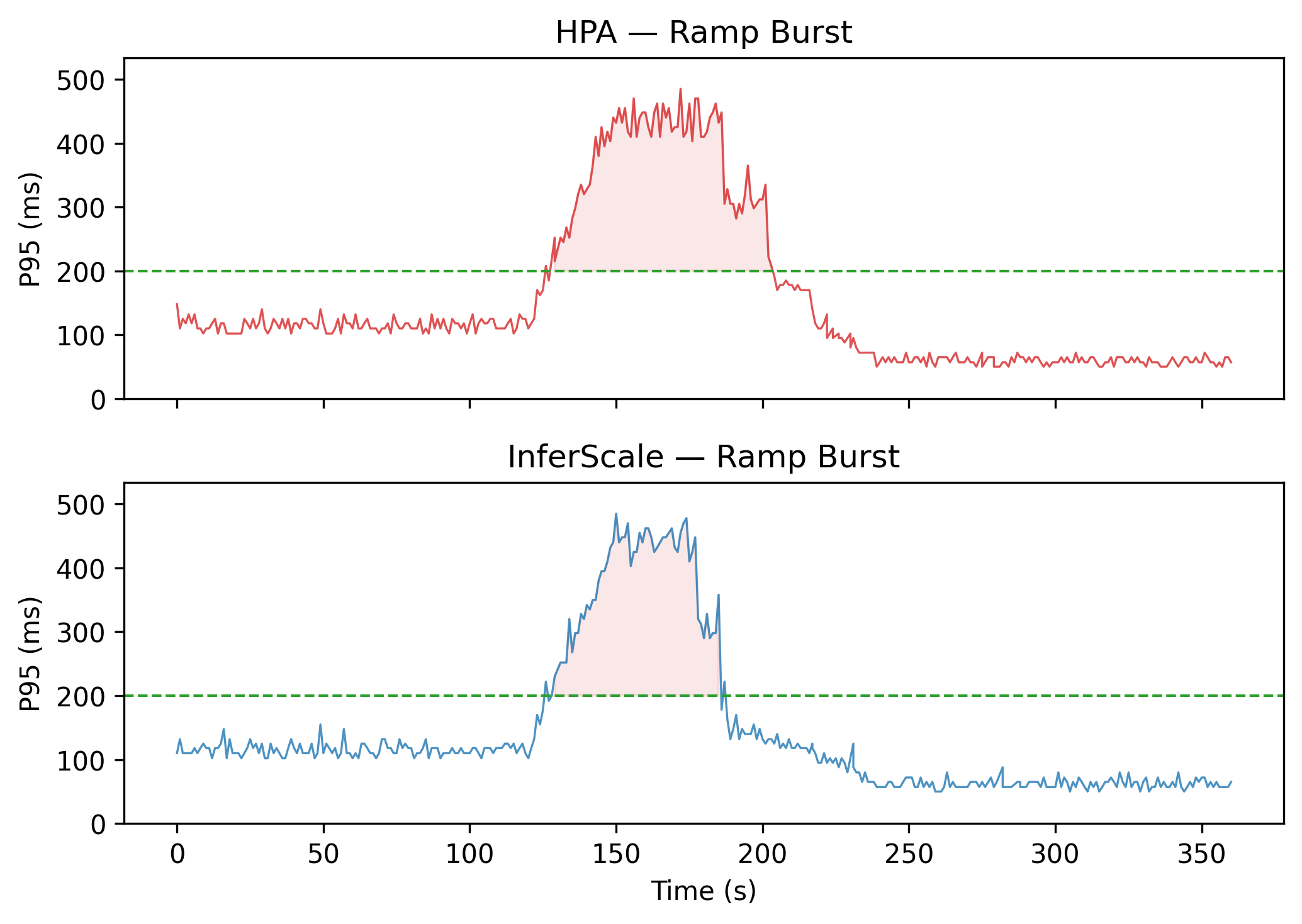}
    \caption{Ramp-burst latency traces for CPU HPA and InferScale. The burst window is short relative to pod startup time, so waiting for clear reactive evidence leaves the controller behind the workload. InferScale moves earlier and keeps the service closer to the target operating region.}
    \label{fig:rampburst}
\end{figure}

Figure~\ref{fig:rampburst} makes the same point visually. CPU HPA spends most of the burst above the SLO line and only begins to recover after load is already falling. InferScale still violates the SLO near the peak, but it spends much more of the burst in the recoverable region rather than in prolonged distress.

What matters here is not only the peak value, but how long the system stays in trouble once it falls behind. In the CPU-HPA run, the service enters overload and more or less stays there until demand is already on the way down. In the InferScale run, the service still gets stretched near the peak, but it does not spend the whole burst in that state. From an operator's perspective, those two traces feel very different even before looking at the aggregate SLO numbers: one is a long period of visible service degradation, the other is a shorter period of strain followed by recovery.

There is also an operational asymmetry. Under CPU HPA, the load generator timed out during sustained overload. External HPA and InferScale both complete with full coverage. The missing CPU-HPA data are therefore not just a measurement annoyance---they are part of the operational outcome of allowing the service to enter prolonged distress. The aggregate CPU-HPA ramp value in Table~\ref{tab:realmodel} is computed over available samples and includes recovery-period observations, whereas Table~\ref{tab:ramp-phases} focuses on the four ramp phases where the controller falls behind.

\begin{table}[t]
\centering
\caption{Ramp-burst phase analysis (real model, CPU HPA vs InferScale). Latency in ms.}
\label{tab:ramp-phases}
\small
\begin{tabular}{@{}lcccc@{}}
\toprule
\textbf{System} & \textbf{0--60s} & \textbf{60--120s} & \textbf{120--180s} & \textbf{180--240s} \\
 & \textbf{2 QPS} & \textbf{6 QPS} & \textbf{12 QPS} & \textbf{18 QPS} \\
\midrule
CPU HPA SLO\% & 100\% & 0\% & 0\% & 0\% \\
IS SLO\% & 100\% & \textbf{91\%} & \textbf{66\%} & 8\% \\
\midrule
CPU HPA P95 & 628 & 3600 & 3606 & 5380 \\
IS P95 & 306 & \textbf{778} & \textbf{1395} & 2165 \\
\bottomrule
\end{tabular}
\end{table}

Table~\ref{tab:ramp-phases} breaks the ramp result down by phase for the operational-default baseline, CPU HPA, versus InferScale. At 6\,QPS, InferScale is already at six replicas and largely within SLO while CPU HPA is in complete violation with one replica. At 12\,QPS, InferScale maintains 66\% compliance near the operating boundary; CPU HPA still shows zero. Even at 18\,QPS---which we observe to overload seven pods at the target SLO in this setup---InferScale degrades more gracefully (P95 2165\,ms vs 5380\,ms). The stronger external HPA baseline in Table~\ref{tab:realmodel} narrows the gap, but this detailed phase view shows why early scaling still matters. The absolute ramp SLO remains below 100\% not because the controller stops helping, but because the combination of 21-second cold start and bounded cluster capacity leaves part of the burst unsupportable at the target SLO.

The diurnal pattern tells a similar story: during the recovery phase (180--270s, 6\,QPS), InferScale achieves 93\% SLO compliance versus CPU HPA's 78\%, because it already has enough replicas from its earlier scaling response.

\textbf{Scaling timeline.} InferScale's first scale-up occurs within roughly the first 15 seconds (by the fourth 5-second sample), reaching six replicas before the heavy-load phase begins. CPU HPA first detects sustained pressure and issues a scale-up decision at roughly 168 seconds; by 180 seconds, the effect is visible in the observed replica count. This more-than-150-second head start is the core mechanism behind the aggregate numbers.

\begin{table}[t]
\centering
\caption{Scaling timeline under ramp burst (real model, CPU HPA vs InferScale).}
\label{tab:timeline}
\small
\begin{tabular}{@{}lcc@{}}
\toprule
\textbf{Time} & \textbf{CPU HPA replicas} & \textbf{InferScale replicas} \\
\midrule
0\,s & 1 & 1 \\
30\,s & 1 & \textbf{3} (early scale-up) \\
60\,s & 1 & \textbf{6} \\
120\,s & 1 & 6 \\
180\,s & \textbf{2} (first scale) & \textbf{7} \\
330\,s & \textbf{3} (too late) & 6 \\
\bottomrule
\end{tabular}
\end{table}

Table~\ref{tab:timeline} highlights the timing gap between CPU HPA and InferScale. By the time CPU HPA creates its second replica at 180 seconds, InferScale is already at maximum capacity with seven replicas. CPU HPA reaches three replicas only at 330 seconds, when load has already dropped to 2\,QPS.

\subsection{Q2: Multi-tenant contention}

Three tenants (Gold/Silver/Bronze) share a cluster with approximately 10.5 allocatable CPUs while total demand reaches 240\,QPS during the stress phase.

\begin{table}[t]
\centering
\caption{Multi-tenant contention (controlled service, mean $\pm$ std, $n$=5).}
\label{tab:multitenant}
\small
\begin{tabular}{@{}lcc@{}}
\toprule
\textbf{Metric} & \textbf{External HPA} & \textbf{InferScale} \\
\midrule
Gold SLO attainment & 89.9 $\pm$ 16.1\% & \textbf{97.9 $\pm$ 1.2\%} \\
Silver SLO attainment & 84.3 $\pm$ 22.1\% & 89.3 $\pm$ 23.8\% \\
Bronze SLO attainment & 94.7 $\pm$ 11.8\% & 94.7 $\pm$ 11.9\% \\
\midrule
Gold median P95 & 142 & \textbf{130} \\
Silver median P95 & \textbf{153} & 191 \\
Bronze median P95 & \textbf{134} & 179 \\
\bottomrule
\end{tabular}
\end{table}

The key finding is not that all tenants become uniformly stable. Rather, InferScale makes contention outcomes more \emph{structured}. Under the external-metric HPA baseline, the victim tenant changes from trial to trial: in one run Silver collapses while Gold remains healthy; in another, Gold is the one that fails. Under InferScale, the Gold-tier tenant is consistently protected, with 97.9\% $\pm$ 1.2\% SLO attainment and a worst case of 96.4\%, which still exceeds the external-HPA Gold mean of 89.9\%.

\begin{table}[t]
\centering
\caption{Per-trial multi-tenant SLO attainment (\%, $n$=5).}
\label{tab:pertrial}
\tiny
\begin{tabular}{@{}lccccc|ccccc@{}}
\toprule
 & \multicolumn{5}{c|}{\textbf{Ext. HPA}} & \multicolumn{5}{c}{\textbf{InferScale}} \\
\textbf{Tenant} & t1 & t2 & t3 & t4 & t5 & t1 & t2 & t3 & t4 & t5 \\
\midrule
Gold & 98.6 & 100 & 61.8 & 98.3 & 90.5 & 97.3 & 97.4 & 99.0 & 96.4 & 99.1 \\
Silver & 68.3 & 100 & 100 & 100 & 53.3 & 100 & 100 & 100 & 99.9 & 46.7 \\
Bronze & 73.5 & 100 & 100 & 100 & 100 & 100 & 100 & 100 & 73.3 & 100 \\
\bottomrule
\end{tabular}
\end{table}

Table~\ref{tab:pertrial} makes the distinction concrete. The external-metric HPA baseline produces \emph{trial-dependent victim selection}: trial~1 hurts Silver, trial~3 hurts Gold, and trial~5 again hurts Silver. InferScale does not remove contention effects for all tenants. Instead, it makes the degradation pattern more policy-aligned: Gold remains consistently protected, while lower-priority tenants absorb more of the contention when resources are tight.

This tradeoff is visible in the replica distributions as well. Under the external-metric HPA baseline, whichever tenant happens to scale first can capture most of the available headroom. Under InferScale, allocations are still differentiated, but the controller uses that differentiation to preserve Gold-tier SLOs rather than allowing the losing tenant to vary unpredictably from trial to trial. The benefit is therefore not ``universal stability,'' but predictable priority enforcement under scarcity. This is not a free win: under scarcity, preserving Gold-tier SLOs can lead to worse latency for lower-priority tenants, which is visible in the Silver and Bronze median P95 values in Table~\ref{tab:multitenant}. The Silver t5 and Bronze t4 outliers illustrate the limit of autoscaling without scheduler coordination: once contention becomes acute, a scheduler-blind controller can enforce policy priorities more reliably than it can eliminate lower-tier degradation.

\begin{figure}[t]
    \centering
    \includegraphics[width=0.92\linewidth]{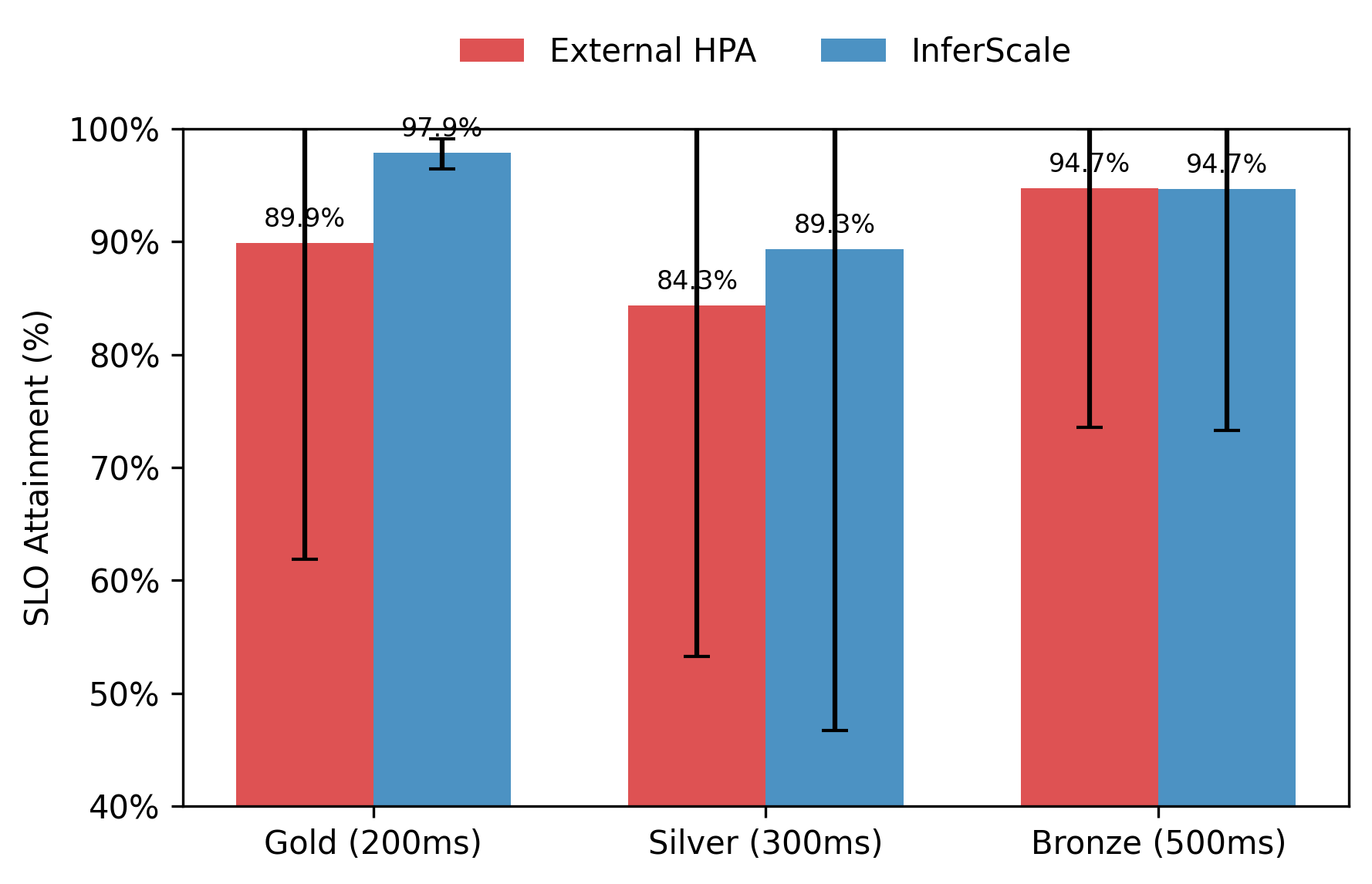}
    \caption{Per-tenant SLO attainment across repeated trials ($n$=5). Bars show means and whiskers show the observed min--max range, bounded by the feasible 0--100\% SLO interval. The external-metric HPA baseline yields trial-dependent victim selection, whereas InferScale consistently protects the Gold-tier tenant and makes degradation more priority-aligned under scarcity.}
    \label{fig:multitenant}
\end{figure}

Figure~\ref{fig:multitenant} shows the same pattern visually. Averages hide part of the change: the cluster does not suddenly have enough capacity for everyone. What changes is who pays when capacity runs out. With the reactive baseline, that cost lands in different places in different runs. With InferScale, the cost lands more often where the policy says it should land. For a platform operator, that is the practical difference between an unstable service and a stable policy.

\subsection{Q3: Ablation}

\begin{table}[t]
\centering
\caption{Ablation: single-service real model (P95 $<$ 1500\,ms). Ramp results are repeated means ($n$=3); diurnal results are representative single-run values.}
\label{tab:ablation-single}
\small
\begin{tabular}{@{}lccc@{}}
\toprule
\textbf{Variant} & \textbf{Scenario} & \textbf{SLO\%} & \textbf{$\Delta$} \\
\midrule
Full InferScale & Ramp & \textbf{65.3 $\pm$ 9.0\%} & --- \\
No-Predict & Ramp & 59.5 $\pm$ 1.3\% & $-$5.8\,pp \\
No-FastPath & Ramp & 57.3 $\pm$ 1.8\% & $-$8.0\,pp \\
\midrule
Full InferScale & Diurnal & \textbf{80.3\%} & --- \\
No-Predict & Diurnal & 74.0\% & $-$6.3\,pp \\
\bottomrule
\end{tabular}
\end{table}

\begin{table}[t]
\centering
\caption{Ablation: priority removal under multi-tenant contention (single trial, exploratory directional evidence).}
\label{tab:ablation-priority}
\small
\begin{tabular}{@{}lcc@{}}
\toprule
\textbf{Variant} & \textbf{Silver SLO\%} & \textbf{$\Delta$} \\
\midrule
Full InferScale & \textbf{100\%} & --- \\
No-Priority & 73\% & $-$27\,pp \\
\bottomrule
\end{tabular}
\end{table}

Table~\ref{tab:ablation-single} shows that prediction and fast-path handling are complementary: in repeated ramp runs, removing prediction reduces SLO by 5.8\,pp and removing fast-path handling reduces it by 8.0\,pp. The predictor captures the upward trend, whereas fast-path reacts at the rising edge. Taken together, these results suggest that the practical gain does not come from prediction alone. Under startup-sensitive ramp bursts, exposing and acting on rising-demand signal early is at least as important as the forecasting component itself for preserving SLOs. The single-run diurnal result suggests a smaller effect under smooth load ($-$6.3\,pp), which matches the broader observation that stronger reactive control catches up when demand changes more gradually. Table~\ref{tab:ablation-priority} reports the priority-removal ablation. Priority control matters under contention, but we treat this result as exploratory rather than definitive: without priority differentiation, Silver drops from 100\% to 73\%, which is consistent with the role of structured priority enforcement but not yet a stable estimate.

\subsection{Overhead}

We do not claim a comprehensive overhead study. In our setting, the dominant costs are pod startup (about 21\,s in the real-model study), readiness delay, and request latency under overload rather than the autoscaler logic itself. InferScale follows the standard operator pattern, queries a small set of Prometheus metrics, and patches Deployment replicas through the Kubernetes API. It does not modify the serving runtime or replace the scheduler. In that sense, the control-path cost is small relative to the startup and serving costs that dominate the evaluated scenarios.

\section{Discussion and Limitations}

InferScale helps when there is enough trend signal to move before the burst fully lands. A pure step increase with no precursor is still hard because cold start is a physical limit.

The paper also has a clear boundary. The real-model results are single-service, while the multi-tenant results use a controlled service. We did explore a real-model multi-tenant ONNX setting, but it did not strengthen the paper. Cold start remained dominant, and the per-pod CPU setting was not tight enough for scarcity-based prioritization to show up reliably. We therefore keep the two evidence lines separate.

The updated multi-tenant results are useful partly because they are not a perfect win. InferScale does not make every tenant stable. Under the external-metric HPA baseline, the losing tenant changes from run to run. Under InferScale, Gold stays protected much more reliably, and the cost of that decision shows up on lower-priority tenants when the cluster is tight. This matters operationally: for a platform team, a predictable policy tradeoff is easier to reason about than random victim selection.

There is also a measurement caveat in the real-model study. The CPU-HPA runs contain fewer data points because the load generator timed out under sustained overload. We therefore report SLO attainment over the available windows rather than over the nominal full schedule. This is not ideal, but the failures are part of the operational outcome rather than just a logging artifact.

The current results come from one model family, two single-service traffic patterns, and one repeatable shared-cluster setup. That is enough to support the paper's main takeaways, but not universal gains across all inference workloads.

Finally, this is an autoscaling paper, not a scheduling paper. The controller decides when and how much to scale; it does not decide where pods land. That choice keeps the design simple, but it also explains some of the lower-tier outliers under acute contention. If tighter guarantees are needed for lower-priority tenants as well, scheduler coordination is the next place to look.

\textbf{Generalizability.} The current evaluation uses CPU-based inference with ONNX Runtime. GPU-backed inference would change startup and capacity dynamics, and evaluating that regime is outside the scope of this paper. For LLM serving, capacity should likely be expressed in prefill/decode token throughput, KV-cache pressure, batching state, and GPU memory headroom rather than only QPS per replica. Startup also includes image pull, weight loading, GPU memory allocation, and sometimes cache warmup. The same underlying problem still matters there, but the capacity model, metrics, and guardrail settings would need workload-specific recalibration. InferScale also targets online inference rather than training jobs. Elastic training has different objectives, such as throughput, job completion time, checkpoint cost, and gang scheduling constraints; we do not claim that the current controller directly generalizes to that setting.

\section{Related Work}

\textbf{Inference serving systems.} Clipper~\cite{clipper} established the case for low-latency prediction serving. INFaaS~\cite{infaas}, Cocktail~\cite{cocktail}, and InferLine~\cite{inferline} study adaptation, provisioning, and scaling for inference pipelines under changing demand. These systems reason about model selection, serving pipelines, and model-serving resources inside or close to the serving stack. InferScale is complementary: it operates at the Kubernetes replica-control layer and keeps the serving runtime unchanged. This makes the design less specialized than a serving-system scheduler, but easier to attach to ordinary Kubernetes deployments that already export request, latency, and queue metrics.

\textbf{Kubernetes and serverless autoscaling.} Kubernetes HPA~\cite{k8shpa} provides the default replica-control mechanism used by many deployments. KEDA~\cite{keda} and Knative~\cite{knative} broaden the trigger set and support event-driven or scale-to-zero operation. These systems are valuable operationally, and our external-metric HPA baseline uses inference-specific queue depth to avoid a weak CPU-only comparison. However, they remain largely reactive: once the metric crosses the target, pod startup and readiness still determine when capacity arrives. InferScale keeps the same Kubernetes control surface but changes the control policy around startup-sensitive inference.

\textbf{Predictive autoscaling.} Predictive autoscaling has a long history in cloud systems, and MArk~\cite{mark} is especially relevant because it uses forecasting to reduce reaction lag for machine-learning inference. InferScale shares that motivation but differs in three ways. First, it operates directly at the Kubernetes replica-control layer rather than assuming a specialized serving environment. Second, prediction is only one input to an SLO-state controller with fast-path spike handling. Third, the paper studies shared-cluster priority handling, where the question is not only whether the average service scales earlier but whether contention produces predictable policy outcomes.

\textbf{Resource management and cloud control.} Borg~\cite{borg}, Autopilot~\cite{autopilot}, and FIRM~\cite{firm} represent the broader cluster-management and production autoscaling context in which our problem sits. These systems show that safe automation, resource isolation, and SLO-oriented control matter at production scale. InferScale addresses a narrower problem at the Kubernetes replica-control layer, but the same lesson applies: useful autoscaling should be judged by operational outcomes, not only by prediction accuracy.

\textbf{Modern AI serving.} vLLM~\cite{vllm} and Llumnix~\cite{llumnix} focus on runtime-level optimizations such as PagedAttention, memory management, and request migration for LLM serving. Those mechanisms change the capacity surface exposed to an autoscaler: token throughput, batching efficiency, and KV-cache pressure may matter more than CPU utilization or raw QPS. InferScale does not replace such runtime mechanisms. Instead, it could consume their exported metrics in the same way it consumes queue depth and P95 latency in this paper.

\section{Conclusion}

We presented InferScale, a Kubernetes-native predictive autoscaler for AI inference that combines lightweight trend prediction, SLO-state control, and tenant-priority headroom without replacing the default scheduler. In repeated ramp-burst runs, InferScale lifts SLO compliance from 47.8\% $\pm$ 1.5\% under a stronger external-metric HPA to 65.3\% $\pm$ 9.0\%. Under repeated multi-tenant contention, the main benefit is not uniform stability but more reliable Gold-tier protection and a more policy-aligned degradation pattern.

The main limitation is still evidence composition: the real-model results are single-service while the multi-tenant results use a controlled service. The clearest next steps are real multi-tenant model serving under active scarcity and tighter coordination between autoscaling and scheduling.

\IEEEtriggeratref{8}
\bibliographystyle{IEEEtran}
\bibliography{references}

\end{document}

%% file: fig_inferscale_arch.tex
\begin{figure}[t]
\centering
\setlength{\fboxsep}{6pt}
\small
\fbox{%
\begin{minipage}{0.92\linewidth}
\centering
\textbf{InferScale control architecture}\\[0.4em]

\fbox{\parbox{0.20\linewidth}{\centering
\textbf{Autoscaler CR}\\
SLO target\\
Tenant class\\
Replica bounds\\
Forecast window\\
Cooldown
}}
\hfill
\fbox{\parbox{0.20\linewidth}{\centering
\textbf{Prometheus}\\
QPS\\
P95 latency\\
Queue depth\\
Ready replicas
}}
\hfill
\fbox{\parbox{0.20\linewidth}{\centering
\textbf{Kubernetes API}\\
Read/patch Deployment\\
Replica updates
}}\\[0.6em]

$\Downarrow$ config \hfill $\Downarrow$ metrics \hfill $\Downarrow$ actuation \\[0.4em]

\fbox{\parbox{0.84\linewidth}{\centering
\textbf{InferScale Controller (5s reconcile loop)}\\[0.2em]
Collect metrics $\rightarrow$ predict demand $\rightarrow$ estimate capacity\\
$\rightarrow$ evaluate SLO state $\rightarrow$ guarded replica decision
}}\\[0.6em]

$\Downarrow$ desired / safe replicas\\[0.4em]

\fbox{\parbox{0.32\linewidth}{\centering
\textbf{Target Deployment / Pods}\\
ONNX Runtime server\\
Ready after download + init + warmup\\
Exports serving metrics
}}
\end{minipage}%
}
\caption{InferScale architecture. The custom resource provides policy, Prometheus provides serving metrics, and the controller updates Deployment replicas through the Kubernetes API.}
\label{fig:arch}
\end{figure}